\documentclass[12pt]{article}
\usepackage{graphicx,multirow}

\newcommand\pubnumber{}
\newcommand\pubdate{\today}

\input babarsym

\newcommand {\Bxclnu} {\ensuremath{B \rightarrow X_c \ell \nu}\xspace}
\newcommand {\Bxulnu} {\ensuremath{B \rightarrow X_u \ell \nu}\xspace}

\newcommand {\mx}     {\ensuremath{M_{X}}\xspace}
\newcommand {\Pplus}  {\ensuremath{P_{+}}\xspace}

\newcommand {\Q}      {\ensuremath{q^{2}}\xspace}

\newcommand {\mX}     {\ensuremath{M_{X}}\xspace}

\newcommand {\beq}    {\begin{equation}}
\newcommand {\beqa}   {\begin{eqnarray}}
\newcommand {\beqn}   {\begin{eqnarray}}
\newcommand {\eeq}    {\end{equation}}
\newcommand {\eeqa}   {\end{eqnarray}}
\newcommand {\eeqn}   {\end{eqnarray}}
\makeatletter
\def\slash#1{{\mathpalette\c@ncel{#1}}} 
\makeatother
\newcommand{\Dslash}{\slash D}

\newcommand{\gevccsq}{\ensuremath{{\mathrm{\,Ge\kern -0.1em V^2\!/}c^4}}\xspace}

\def\Ferrara{Istituto Nazionale di Fisica Nucleare\\
Sezione di Ferrara, I-44122 Ferrara, ITALY \\
\vskip 2cm
on behalf of the Babar and Belle Collaborations}

\def\Title#1{\begin{center} {\Large #1 } \end{center}}
\def\Author#1{\begin{center}{ \sc #1} \end{center}}
\def\Address#1{\begin{center}{ \it #1} \end{center}}

\newcommand\pubblock{\rightline{\begin{tabular}{l} \pubnumber\\
         \pubdate  \end{tabular}}}
\newenvironment{Abstract}{\begin{quotation}  }{\end{quotation}}
\newenvironment{Presented}{\begin{quotation} \begin{center} 
             PROCEEDINGS OF CKM2010\end{center}\bigskip 
      \begin{center}\begin{large}}{\end{large}\end{center} \end{quotation}}

\def\beq{\begin{equation}}
\def\eeq#1{\label{#1}\end{equation}}
\def\eeqn{\end{equation}}

\def\beqa{\begin{eqnarray}}
\def\eeqa#1{\label{#1}\end{eqnarray}}
\def\eeqan{\end{eqnarray}}

\let\bar=\overbar

\def\Dslash{\not{\hbox{\kern-4pt $D$}}}
\def\dslash{\not{\hbox{\kern-2pt $\del$}}}

\def\BR{\mbox{\rm BR}}

\def\msb{{\bar{\ssstyle M \kern -1pt S}}}

\begin{document}
\begin{titlepage}
\pubblock

\vfill
\Title{Measurements of Inclusive ${B \rightarrow X_u \ell \nu}\xspace$ Decays}
\vfill
\Author{Concezio Bozzi}
\Address{\Ferrara}
\vfill
\begin{Abstract}
Recent results on inclusive charmless semileptonic decays of B mesons are 
reviewed. Emphasis is given to measurements on the recoil of fully reconstructed B 
mesons, which allow to exploit several regions of phase space. Preliminary averages of  
the CKM  matrix element $|V_{ub}|$ from the Heavy Flavour Working Group are shown, 
using four different theoretical calculations. 
\end{Abstract}
\vfill
\begin{Presented}
the 6th International Workshop on the CKM Unitarity Triangle, 
University of Warwick, UK, 6-10 September 2010 
\end{Presented}
\vfill
\end{titlepage}
\def\thefootnote{\fnsymbol{footnote}}
\setcounter{footnote}{0}

\section{Introduction}
Measurements of inclusive charmless semileptonic decays of B mesons, \Bxulnu, are directly 
related to the CKM matrix element $|V_{ub}|$. The theoretical description~\cite{einan} of the 
hadronic current involved in these decays, relying on the Operator Product Expansion (OPE) 
technique, allows the determination of \Vub\ from the total decay 
rate with a small uncertainty. 
However, in order to suppress background from semileptonic decays with charm, \Bxclnu, measurements 
of partial branching fraction are performed in restricted kinematic regions. 
Unfortunately, OPE breaks down in some of these regions, and the theoretical uncertainty increases significantly. 
Contributions due to weak annihilation also play a role in part of the kinematic regions, 
and need to be carefully assessed. 
In short, theory and backgound subtraction give conflicting requirements, and a 
trade-off must be found. Due to the improved knowledge of \Bxclnu
transitions and to the abundant data samples collected at the B factories, recent results 
based on phase space regions which are increasingly larger allow for an improved precision 
in the $|V_{ub}|$ determinations. 

Experimental measurements of charmless semileptonic decays are 
reviewed in Section~\ref{sec:PBF}, with emphasis on  
new preliminary results from Babar. 
Preliminary \Vub\ averages from the Heavy Flavour Averaging Group by using the available 
theory calculations are presented in Section~\ref{sec:vub}. 
Conclusions are given in Section~\ref{sec:concl}.

\section{Measurements of Partial Branching Fractions}
\label{sec:PBF}
Measurements 
in the endpoint region of the lepton momentum spectrum 
are conceptually simple, being based on the identification of an high 
momentum lepton only. However, the kinematic region selected by the 
high lepton momentum requirement to suppress charmed background suffers from 
sizeable theoretical uncertainties. All experimental efforts  
\cite{endp} have been 
devoted to reducing the lepton momentum cut as low as allowed by background knowledge. 
Signal-to-background ratios (S/B) of the order of 1/10 have 
been achieved, with signal efficiencies at the 30\% level 
or less. An improved analysis \cite{shmax}, based on the measurement of 
missing energy to estimate the maximum kinematically allowed hadronic mass squared,
$s_h^{max}$, resulted in S/B of about 1/2. 
A summary of the available endpoint measurements is given in 
Table~\ref{tab:endpoint}. 

\begin{table}[t]
\begin{center}
\begin{tabular}{ll|ccc}  
\multicolumn{2}{c|}{Experiment} & ${\cal{L}}(fb^{-1})$ & $E_{\ell}$ (GeV) & $\Delta{\cal{B}}(10^{-4})$ \\ \hline
E1 & Babar	& 81.4	& 2.0--2.6 & $5.72\pm 0.41\pm 0.65$ \\
E2 & Belle	& 27.0	& 1.9--2.6 & $8.5\pm 0.4\pm 1.5$ \\
E3 & CLEO	& 9.13	& 2.2--2.6 & $2.30\pm 0.15\pm 0.35$ \\ \hline
E4 & Babar ($s_h^{max}<$3.5 GeV$^2$)& 81.4 & 2.0--2.6 & $4.41\pm 0.42\pm 0.42$ \\ 
\hline
\end{tabular}
\caption{Summary of endpoint analyses, labeled E1--E4 in the following.}
\label{tab:endpoint}
\end{center}
\end{table}
\begin{table}[t]
\begin{center}
\begin{tabular}{lcc|cc} 
     \multicolumn{3}{c|}{Kinematic Region} & Signal Yield & $\Delta {\cal{B}}(\Bxulnu)\ (10^{-3})$ \\ \hline
Babar & R1 & $\mX < 1.55\gevc$ 		   & $1033 \pm 73$  &  $1.08 \pm 0.08 \pm 0.06$  \\
      & R2 & $\mX < 1.70\gevc$		   & $1089 \pm 82$  &  $1.15 \pm 0.10 \pm 0.08$  \\
      & R3 & $\Pplus < 0.66\gev$	   & $ 902 \pm 80$  &  $0.98 \pm 0.09 \pm 0.08$  \\
      & R4 & R2 and $ \Q>8\gevcc$          & $ 665 \pm 53$  &  $0.68 \pm 0.06 \pm 0.04$  \\
      & R5 & $p_\ell^*>1\gevc,~(\mx, \Q)$ fit  & $1441 \pm102$  &  $1.80 \pm 0.13 \pm 0.15$  \\
      & R6 & $p_\ell^*>1.3\gevc$	   & $ 562 \pm 55$  &  $0.76 \pm 0.08 \pm 0.07$  \\ \hline
Belle & R5 & $p_\ell^*>1\gevc,~(\mx, \Q)$ fit  & $1032 \pm 91$  &  $1.96 \pm 0.17 \pm 0.16$  \\
\hline
\end{tabular}
\caption{Summary of signal yields and partial branching fractions in 
six kinematic regions (labeled R1--R6 in the following). 
Unless otherwise noted, the lepton momentum in the center-of-mass frame is required to be $p_\ell^*>1\gevc$. 
The uncertainty on the yields is statistical only.}
\label{tab:inputdeltaB}
\end{center}
\end{table}

The {\em{recoil technique}} aims at fully reconstructing one of the two $B$ mesons ($B_{reco}$) 
from the $\Upsilon (4S)$ decay in a fully hadronic decay, which allows to determine completely 
the decay kinematics of the other $B$ ($B_{recoil}$). 
It is therefore possible to access relevant kinematic variables, such as 
the invariant mass of the hadronic system, $m_X$, the light-cone momentum component 
$P_+=E_X-|{\vec{p}}_X|$, and the squared invariant mass of the lepton pair, $q^2$. 
Semileptonic events are identified by 
an high-momentum lepton ($p_{\ell}^* > 1$ GeV) and a missing mass 
consistent with zero. 
Non semileptonic backgrounds are subtracted by studying the distribution of the 
beam-energy substituted mass $m_{ES}$ for the $B_{reco}$ candidates. 
About 1000 hadronic modes are reconstructed, with efficiencies 
at the 0.3\% (0.5\%) level for neutral (charged) B decays. 

Background from \Bxclnu\ is reduced mainly 
by vetoing charged kaons and $K_S$, whose production is highly suppressed in 
signal, and charged and neutral soft pions kinematically compatible with $B \rightarrow D^* \ell \nu$ 
decays.  

Both B Factory experiments published results using the recoil technique 
\cite{breco}. Babar recently released \cite{ichep2010} a 
preliminary update based on the full dataset (426 $fb^{-1}$), which is detailed 
in the following. The integrated luminosity analysed by Belle is 626 $fb^{-1}$. 

Partial rates for charmless semileptonic 
decays have been measured in several phase space regions, defined in 
Table~\ref{tab:inputdeltaB}, as well as for charged and neutral $B$ decays separately. 
The latter is achieved by explicitly requiring the (absolute) $B_{reco}$ charge 
to be one or zero, respectively, after subtacting with Monte Carlo a small fraction of 
events where the $B_{reco}$ charge was not correctly reconstructed. 
In addition to the $m_X$, $P_+$ and ($m_X, q^2$) distributions, the lepton momentum 
$p_{\ell}^*$ was also studied. 
A background-enriched control sample, obtained by reversing the vetoes on kaons and soft pions 
mentioned above, was used to determine the relative contribution due to semileptonic decays 
into P-wave D mesons directly on data; Monte-Carlo simulation was then reweighted accordingly. 
Although the impact on signal yields was almost 
negligible, the fit chisquares improved significantly. 

\begin{figure}[t]
    \begin{centering}
      \includegraphics[width=0.24\textwidth,totalheight=6.5cm]{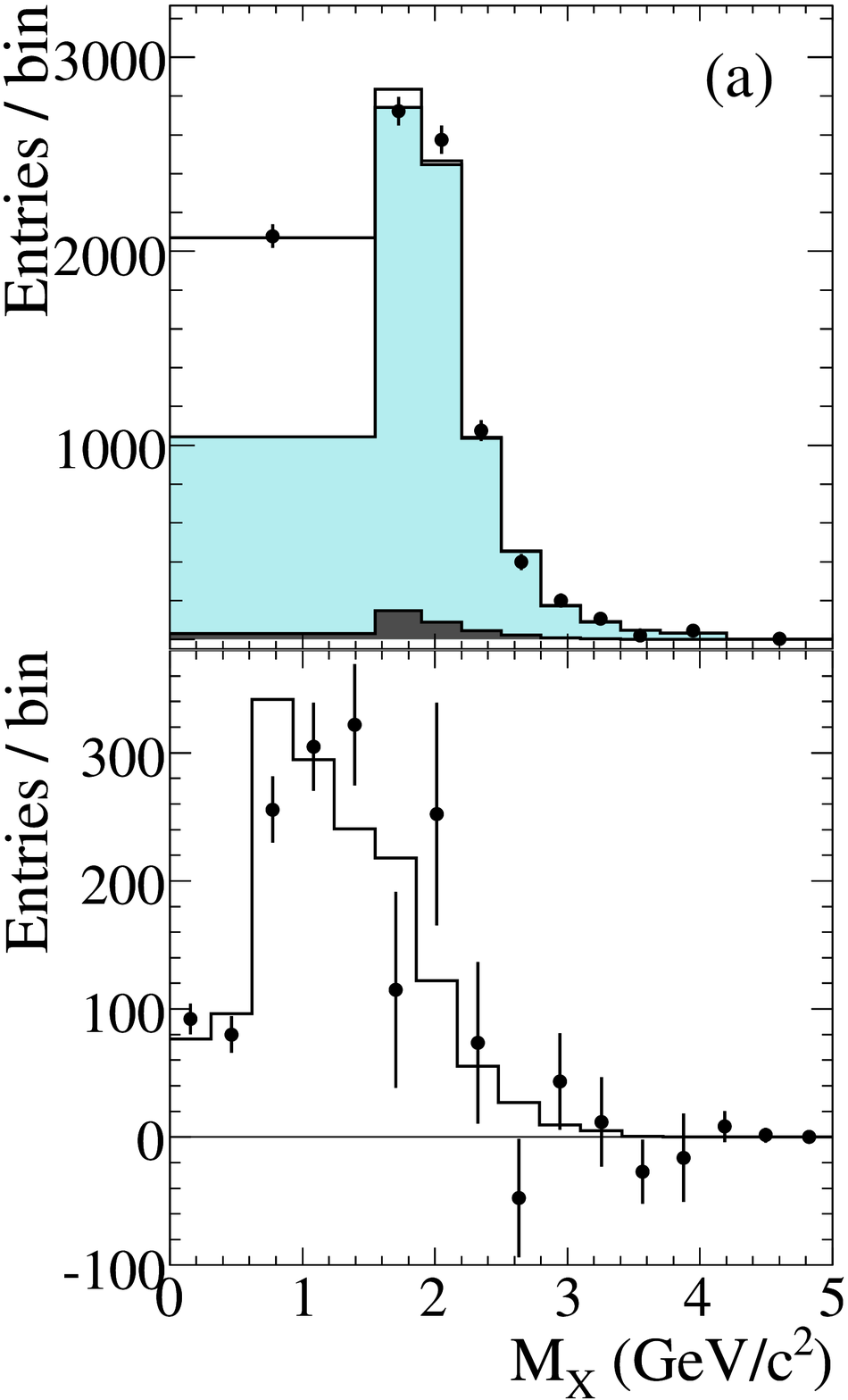}
      \includegraphics[width=0.24\textwidth,totalheight=6.5cm]{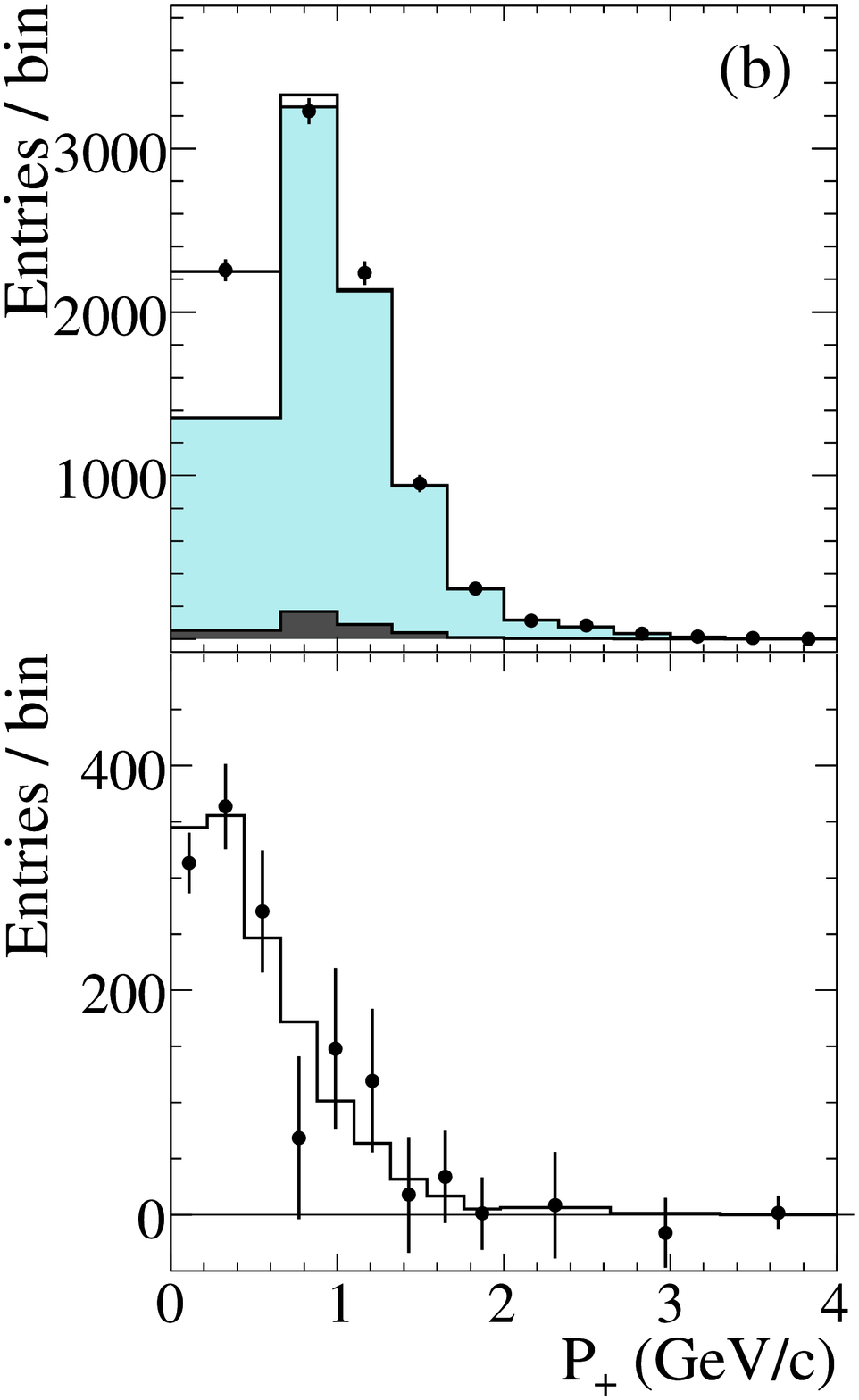}
      \includegraphics[width=0.24\textwidth,totalheight=6.5cm]{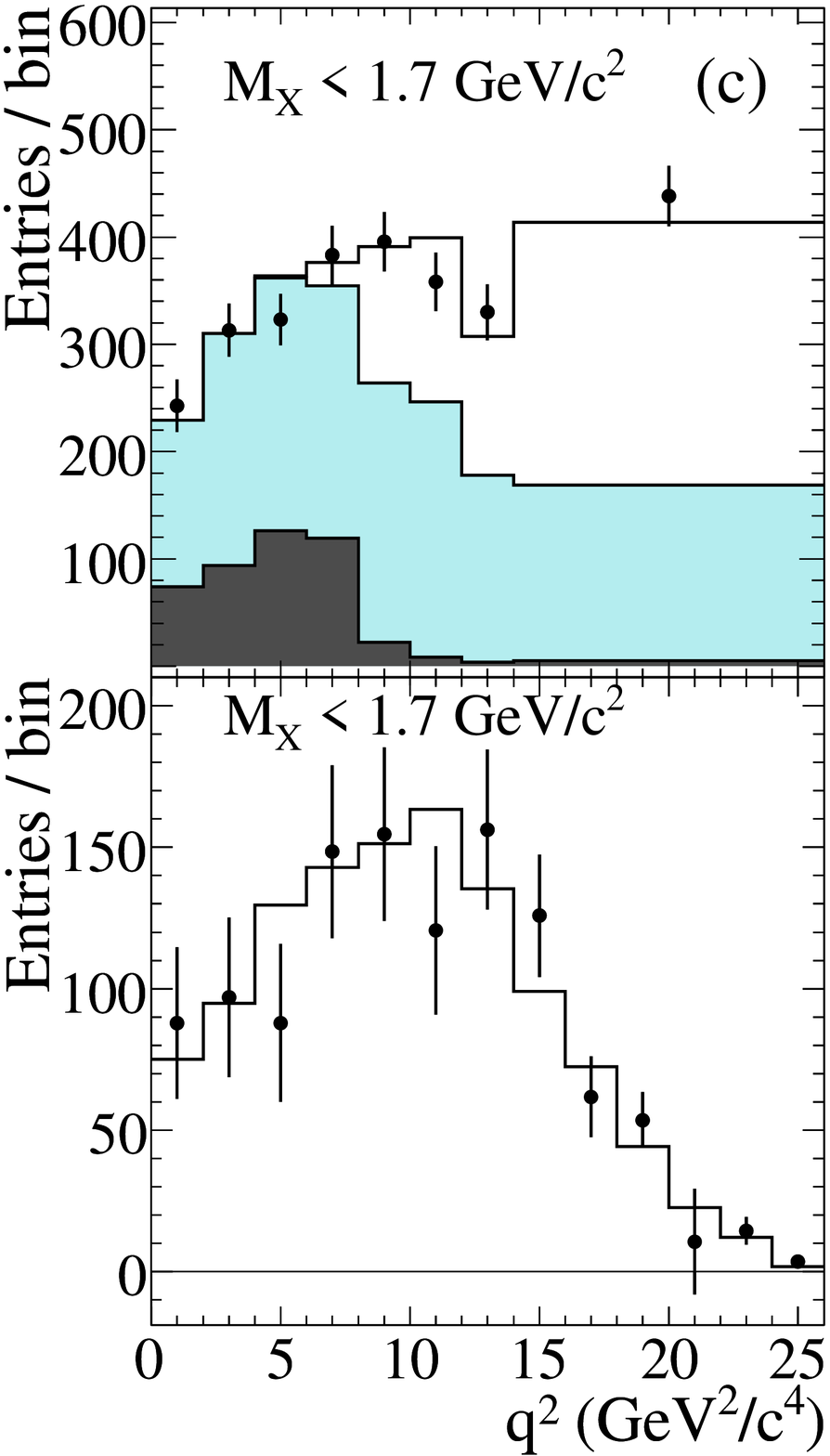}
      \includegraphics[width=0.24\textwidth,totalheight=6.5cm]{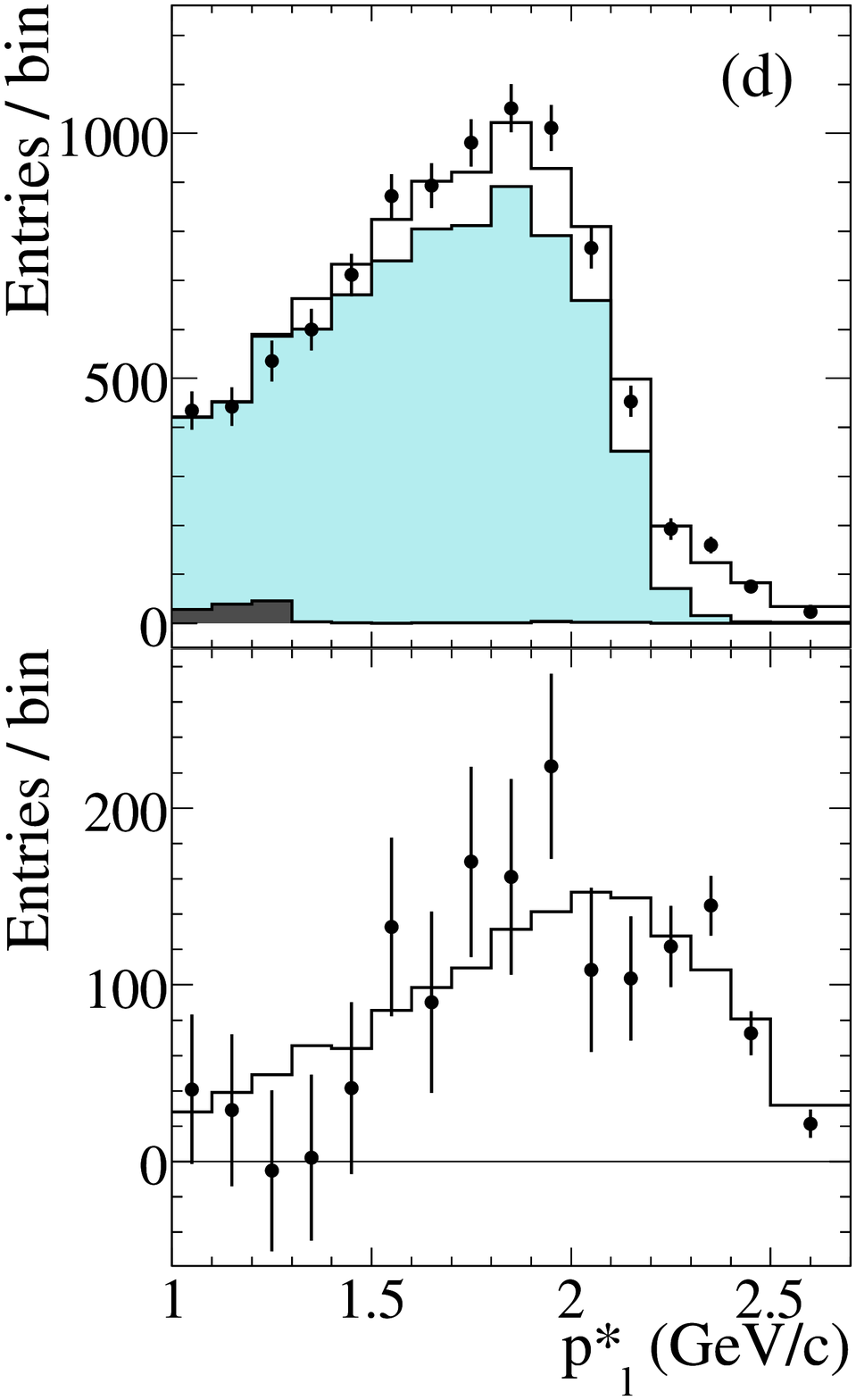}
      \caption{Upper row: \mX\ (a), \Pplus\ (b), \Q\ with $\mX<1.7~\gevcc$ (c) and $p_{\ell}^*$ 
	(d) spectra (data points), measured in Babar data. The result of the fit to the sum of three MC contributions is shown 
        in the histograms: 
	\Bxulnu\ decays generated inside (no shading) and outside (dark shading) the 
	selected kinematic region, and \Bxclnu\ and other background (light shading).  
        Lower row: corresponding spectra for \Bxulnu\ (not corrected for efficiency) 
        after background subtraction.} 
    \label{fig:results}
   \end{centering}
   \end{figure}

The event yields and partial branching fractions
are given in Table~\ref{tab:inputdeltaB}. The distributions of the 
kinematic variables under study, before and after background subtraction, are shown in 
Figure~\ref{fig:results}. Statistical uncertainties range from 7\% to 9\%.

 \begin{table}
   \begin{center}
     \vspace{0.1in}
     \begin{tabular}{l|cccccc|c} 
	&  \multicolumn{6}{|c|}{Babar} 				& Belle \\ 
Source		  & R1	& R2	& R3	& R4	& R5	& R6	& R5 \\ \hline
Statistical error & 7.1	& 8.9	& 8.9	& 8.0	& 7.1	& 8.9	& 8.8 \\
MC statistics 	  & 1.3	& 1.3	& 1.3	& 1.6 	& 1.1	& 1.2	&     \\ 
\hline
Detector-related: & 2.8	& 3.7	& 5.5	& 4.1	& 3.2	& 2.7	& 3.3 \\ 
Fit-related: 	  & 2.7	& 4.9	& 3.2	& 3.2	& 2.1	& 2.5	& 3.6 \\
Signal model:	  & 2.7	& 3.0	& 3.5	& 1.9	& 6.6	& 7.9	& 6.3 \\
Background model: & 2.0	& 2.6	& 3.4	& 2.8	& 2.8	& 2.2	& 1.7 \\ 
\hline
Total systematics: 	& $\mbox{}^{5.3}_{-5.0}$ & $\mbox{}^{6.4}_{-6.2}$   
			& $\mbox{}^{8.0}_{-8.1}$ & $\mbox{}^{6.2}_{-6.2}$    
			& $\mbox{}^{8.5}_{-7.7}$ & $\mbox{}^{9.4}_{-8.7}$ & $\pm 8.1$ \\ 
\hline
Total error:            & $\mbox{}^{9.0}_{-8.8}$ & $\mbox{}^{11.0}_{-10.9}$ 
			& $\mbox{}^{12.0}_{-12.1}$ & $\mbox{}^{10.2}_{-10.3}$  
			& $\mbox{}^{11.1}_{-10.5}$ & $\mbox{}^{12.9}_{-12.4}$ & $\pm 12.0$ \\ \hline
     \end{tabular}
     \caption{Systematic uncertainties (in percent) on the partial branching fractions for the various phase space regions, 
		for the Babar and Belle recoil analyses.}
     \label{tab:systematics}
   \end{center}
 \end{table}
A summary of the systematic uncertainties is given in 
Table~\ref{tab:systematics}, which also shows the corresponding uncertainties from the Belle 
recoil analysis. 
The statistical and experimental systematic uncertainties are 
of the same order. Detector-related uncertainties are dominated by undetected or 
mismeasured particles ({\em{e.g.}} $K_L$ and additional neutrinos) from background. 
Progress on the knowledge of exclusive \Bxclnu\ decays reflects in a 
relatively small uncertainty due to background composition. In the most inclusive 
phase space region (R5), the dominant 
uncertainty is due to the signal model, in particular to the knowledge of heavy quark 
parameters and the branching fraction of exclusive \Bxulnu\ decays, which 
are used in the simulation to determine signal efficiency. 
Total uncertainties range between 9\% and 13\%.

\begin{table}[b]
\begin{center}
\begin{tabular}{l|cc} 
Phase Space Region  & $(R^{+/0}-1)$ & 90\% C.L. on ${\gamma_{WA}}/{\Gamma}$\\
\hline
R1 & -0.020$\pm$0.066$\pm$0.003 & $[-0.13,0.09]$ \\
R2 & 0.071$\pm$0.117$\pm$0.011  & $[-0.12,0.26]$ \\
R4 & 0.042$\pm$0.066$\pm$0.009  & $[-0.07,0.15]$ \\
R5 & 0.109$\pm$0.157$\pm$0.019  & $[-0.15,0.37]$ \\
\hline
\end{tabular}
\end{center}
\caption{\label{tab:WA} 
Results for $(R^{+/0}-1)$ and limits on ${\gamma_{WA}}/{\Gamma}$ 
for the various kinematic regions under study by Babar.}
\end{table}
Measurements of partial rates for charged and neutral 
$B$ mesons allow to determine the relative contribution of weak annihilation to the total rate, 
${\gamma_{WA}}/{\Gamma}$. The resulting 90\% confidence level regions 
are reported in Table \ref{tab:WA}; 
they are in agreement with previous determinations \cite{wa}. 

\section{$|V_{ub}|$ determinations}
\label{sec:vub}
The value of $|V_{ub}|$ is related to the measured partial branching fractions by 
\begin{equation}
\Vub  = \sqrt{\frac{\Delta {\cal{B}}(\Bxulnu)}{\tau_B \cdot \Delta\Gamma_{theory} }}, \\\nonumber
\label{eq:vub}
\end{equation}
\noindent
where the \Bxulnu width according to the applied cuts, $\Delta\Gamma_{theory}$, and its 
uncertainty are determined by four theoretical calculations \cite{BLNP, DGE, GGOU, ADFR}. 
Theoretical uncertainties can be divided in parametric terms, due to uncertainties on heavy quark parameters 
and $\alpha_s$, and non-parametric contributions due, for instance, to higher order terms in 
the heavy quark expansion, weak annihilation, leading and subleading shape functions, 
renormalization scale. 

The procedure for performing the averages is documented in \cite{hfag:end2009}. 
The average B lifetime used is 1.578 ps. 
The input values for the heavy quark parameters have been determined by a global fit in the kinetic scheme, 
translated to the scheme needed by each model, where both $b\rightarrow c \ell \nu$  
and $b \rightarrow s \gamma$ moments are used, giving $m_b(kin) = 4.591 \pm 0.031$ GeV, 
$\mu_{\pi}^2(kin) = 0.454 \pm 0.038$ GeV$^2$, and a correlation of -40.5\%. 

Preliminary averages from the Heavy Flavour Working Group are shown in Table~\ref{tab:VUB}, 
for the four available theoretical calculations. 
All methods give consistent results and comparable uncertainties. 
\begin{table}[tbp]
\begin{center}
\vspace{0.1in}
\footnotesize
\begin{tabular}{llcccc} 
Kin. 	& Expt. & BLNP 		& DGE 		& GGOU 		& ADFR \\ 
region	&       & \cite{BLNP}	& \cite{DGE}	& \cite{GGOU}	& \cite{ADFR} \\ \hline  
\multicolumn{6}{l}{Endpoint analyses} \\
E1 & Babar & $4.35\pm 0.25 ^{+0.31}_{-0.30}$ & $4.15\pm 0.28 ^{+0.28}_{-0.25}$	& $4.17\pm 0.24 ^{+0.20}_{-0.33}$ & $3.98\pm 0.27 ^{+0.24}_{-0.25}$ \\
E2 & Belle & $4.81\pm 0.45 ^{+0.32}_{-0.29}$ & $4.66\pm 0.43 ^{+0.26}_{-0.25}$	& $4.65\pm 0.43 ^{+0.19}_{-0.30}$ & $4.53\pm 0.42 ^{+0.27}_{-0.27}$ \\
E3 & CLEO  & $4.00\pm 0.47 ^{+0.34}_{-0.34}$ & $3.70\pm 0.43 ^{+0.30}_{-0.26}$	& $3.81\pm 0.44 ^{+0.22}_{-0.39}$ & $3.47\pm 0.41 ^{+0.21}_{-0.22}$ \\
E4 & Babar & $4.48\pm 0.30 ^{+0.39}_{-0.37}$ & $4.15\pm 0.28 ^{+0.30}_{-0.30}$	& n.a.				  & $3.87\pm 0.26 ^{+0.24}_{-0.24}$ \\
\hline
Sim.ann.& Belle	& $4.39\pm 0.46 ^{+0.31}_{-0.29}$ & $4.30\pm 0.45 ^{+0.24}_{-0.23}$ & $4.24\pm 0.45 ^{+0.25}_{-0.33}$	& $3.94\pm 0.41 ^{+0.23}_{-0.24}$ \\ 
\hline
\multicolumn{6}{l}{Recoil Analyses} \\
R1 & Babar & $4.03\pm 0.19 ^{+0.28}_{-0.26}$ & $4.23\pm 0.20 ^{+0.22}_{-0.19}$	& $3.96\pm 0.18 ^{+0.24}_{-0.27}$ & $3.86\pm 0.18 ^{+0.24}_{-0.25}$ \\
R2 & Babar & $3.92\pm 0.22 ^{+0.25}_{-0.23}$ & $4.04\pm 0.22 ^{+0.26}_{-0.23}$	& $3.84\pm 0.21 ^{+0.17}_{-0.20}$ & $3.78\pm 0.21 ^{+0.23}_{-0.24}$ \\
R3 & Babar & $3.90\pm 0.24 ^{+0.28}_{-0.26}$ & $3.93\pm 0.24 ^{+0.36}_{-0.29}$	& $3.64\pm 0.22 ^{+0.30}_{-0.30}$ & $3.60\pm 0.22 ^{+0.23}_{-0.24}$ \\
R4 & Babar & $4.22\pm 0.22 ^{+0.30}_{-0.28}$ & $4.10\pm 0.22 ^{+0.23}_{-0.22}$	& $4.07\pm 0.22 ^{+0.24}_{-0.32}$ & $3.78\pm 0.20 ^{+0.23}_{-0.23}$ \\
R5 & Babar & $4.27\pm 0.24 ^{+0.23}_{-0.20}$ & $4.34\pm 0.24 ^{+0.15}_{-0.15}$	& $4.29\pm 0.24 ^{+0.11}_{-0.14}$ & $4.34\pm 0.24 ^{+0.15}_{-0.15}$ \\
R6 & Babar & $4.22\pm 0.27 ^{+0.23}_{-0.21}$ & $4.27\pm 0.27 ^{+0.16}_{-0.16}$	& $4.21\pm 0.27 ^{+0.12}_{-0.16}$ & $4.28\pm 0.27 ^{+0.26}_{-0.25}$ \\
R5 & Belle & $4.45\pm 0.27 ^{+0.24}_{-0.21}$ & $4.53\pm 0.27 ^{+0.15}_{-0.15}$	& $4.47\pm 0.27 ^{+0.11}_{-0.15}$ & $4.55\pm 0.30 ^{+0.27}_{-0.27}$ \\
\hline
\multicolumn{2}{c}{Average} 	& $4.30\pm 0.16 ^{+0.21}_{-0.23}$ & $4.37\pm 0.15 ^{+0.17}_{-0.16}$	
				& $4.30\pm 0.16 ^{+0.13}_{-0.20}$ & $4.05\pm 0.13 ^{+0.24}_{-0.21}$ \\
\multicolumn{2}{c}{$\chi^2/d.o.f.$ (CL)} & 12.2/11 (0.36) & 7.52/11 (0.76) & 12.2/10 (0.27) & 28.2/11 (0.003)\\
\hline
\end{tabular}
\caption{Results for \Vub\ obtained with four theoretical calculations. The uncertainties are experimental ({\em{i.e.}} 
sum of statistical and experimental systematical) and theoretical, respectively.  
The phase space regions are defined in Tables~\ref{tab:endpoint} and \ref{tab:inputdeltaB}.}
\label{tab:VUB}
\end{center}
\end{table}

A recent NNLO calculation \cite{NNLO} of the leading term in the partial rates gives a surprising change 
with respect to the NLO calculation used in the BLNP method, therefore suggesting an underestimate of the 
theoretical uncertainty due to the renormalization matching scale. The corresponding change in $|V_{ub}|$ is 
of the order of 8\%. Similar estimates for the other methods are not available yet. 
This effect has not been taken into account in the current $|V_{ub}|$ determinations. 

\section{Conclusion}
\label{sec:concl}
Several \Vub\ determinations from measurements of inclusive spectra on 
the full data sets collected at the B Factories are now available. 
In the most inclusive kinematic regions, the dominant error is due to 
uncertainties in signal modeling, which directly propagate in the selection efficiency. 
This uncertainty can be reduced by applying stricter kinematic cuts, but theoretical 
uncertainties correspondingly increase. Background modeling dominates endpoint analyses. 
Statistical and systematic uncertainties on the 
partial rates are comparable. Contributions due to weak annihilation are 
now being addressed with data; however, the statistical sensitivity is still beyond theoretical expectations. 

The uncertainty on inclusive $|V_{ub}|$ determinations is at the 6\% level, dominated by parametric errors 
(4\% from a 40 MeV uncertainty on the $b$ quark mass). Determinations by using the four available  
calculations are consistent between each other; all methods give comparable theory uncertainties, and the spread 
among calculations is comparable to theory errors. Recent results on NNLO calculations might hint to an underestimate 
of non-parametric theory uncertainties, which needs to be clarified.


\begin{thebibliography}{99}
\bibitem{einan}
a comprehensive and critical review of the theoretical framework and calculations 
is given by Einan Gardi in these Proceedings.
\end{thebibliography}
\end{document}